\documentclass[aps,prl,twocolumn,nofootinbib,groupedaddress,amsfonts,floatfix]{revtex4}

\usepackage{graphicx}
\newcommand{\Einit}{E_{\mbox{\tiny init}}}
\newcommand{\tinit}{t_{\mbox{\tiny init}}}
\newcommand{\Hinit}{H_{\mbox{\tiny init}}}
\begin{document}

\title{GUT-Scale Primordial Black Holes: Consequences and Constraints}

\author{Richard Anantua${}^1$}
\author{Richard Easther${}^1$}
\author{John T. Giblin, Jr${}^{1,2}$}

\affiliation{${}^1$Department of Physics, Yale University, New Haven CT 06520}   
\affiliation{${}^2$Department of Physics and Astronomy, Bates College, 44 Campus Ave, Lewiston, ME 04240}
\date{\today}

\begin{abstract}
A population of  very light primordial black holes  which  evaporate before nucleosynthesis begins is unconstrained unless   the decaying black holes leave stable relics. We show that gravitons Hawking radiated from these black holes would source a substantial stochastic background of high frequency gravititational waves ($10^{12}$ Hz or more) in the present universe. These black holes may lead to a transient period of matter dominated expansion. In this case the primordial universe could be temporarily dominated by large clusters of ``Hawking stars" and the resulting gravitational wave spectrum is independent of the initial number density of primordial black holes.
 
\end{abstract}


\maketitle

 
Primordial black holes (PBH) produced immediately after the big bang \cite{Hawking:1971ei,Carr:1974nx} can decay via the emission of  Hawking radiation \cite{Hawking:1974sw,Hawking:1974rv}.  The initial PBH population is determined by the primordial perturbation spectrum. Bounds on the PBH population constrain inflation and other early universe scenarios,  which generate this spectrum  \cite{Green:1997sz,Leach:2000ea,Kohri:2007qn,Peiris:2008be}.   These constraints follow from   the lack of evidence for the present-day existence of PBH, and  because decaying black holes disrupt nucleosynthesis, recombination and reionization \cite{Mack:2008nv}.      At formation, a PBH must be smaller than the Hubble horizon, and the amount of material inside the Hubble horizon -- and the maximal mass and lifetime of a PBH -- {\em increases\/} as the universe expands. Very light PBH  decay completely before nucleosynthesis, and are consequently unconstrained. A PBH radiates any and all particles whose rest mass is substantially less than its current temperature, including gravitons.   If  the PBH radiate  massive, long-lived particles one obtains tight bounds on their initial population \cite{MacGibbon:1987my,Barrow:1992hq,Green:1999yh,Lemoine:2000sq,Khlopov:2004tn} but these limits are contingent upon assumptions about particle physics and quantum gravity, and other radiated particles will reach thermal equilibrium, erasing any memory of their origin. However gravitons emitted as the black hole decays cannot equilibriate and will always survive until the present day, producing a stochastic background of gravitational waves. Further, for some parameter choices the early universe has  a transient matter-dominated phase, during which large clusters of PBH can form. In this case  the resulting gravitational wave spectrum is independent of the initial fraction of black holes. 
 
 The mass-fraction of PBH is denoted $\Omega_{BH}$. Initially $\Omega_{BH}   = \beta$, $0<\beta<1$. We assume that the remaining matter consists of radiation.      PBH form if  $\delta \rho/\rho \gtrsim 10^{-2}$ at very short scales, and a number of inflationary models have this property (\cite{GarciaBellido:1996qt}, \cite{Saito:2008jc} and references therein).   Beyond this threshold $\beta$ rapidly approaches unity.  An appreciable gravitational wave background is generated even if $\beta$ is very small, so given a model which predicts the existence of {\em any} small PBH, the signal discussed here is generic.    
 
For simplicity, we assume a PBH  population whose mass is equal to the energy contained inside the Hubble volume at the instant they collapse.  Recalling that $H^2 = 8\pi \rho /3M_p^2$, and defining $\rho = \Einit^4$,  
\begin{equation}
M_{BH}  = \sqrt{\frac{3}{32 \pi}} \frac{M_p^3}{\Einit^2}
\end{equation}
which is the mass contained inside a sphere of radius $1/H$. Including grey body factors, $\Gamma_{sl}$,  a Schwarzschild black hole emits (massless) particles with momentum $k$,  reducing its total energy as
\begin{eqnarray} \label{eqn:kt}
\frac{dE}{dtdk} &=& -\frac{M_{BH}^2 }{2 \pi   M_p^4  } k  \sum_{s,l} \frac{ (2l+1) h(s) \Gamma_{sl}(\frac{k M_{BH}}{M_p^2})}{\exp{\frac{8 \pi M_{BH} k}{M_p^2}} \pm 1}  \\
&=&
-\frac{2 g }{\pi } \frac{M_{BH}^2 }{M_p^4  }   \frac{k^3}{ e^{k/T} -1} \, \\
T &=& \frac{M_p^2}{8\pi M_{BH}} \, .
\end{eqnarray}
where $h(s)$ counts the helicity/polarization states of  a particle with   spin $s$ \cite{Page:1976df,FrolovBK}.  The second line is the pure black body expression and  $g$ is the {\em effective\/} number of (bosonic) degrees of freedom, after grey body corrections. These suppress emission at larger $s$  so $g$ depends on both the  mix of spin-states and total number of light degrees of freedom. For each state with  $s=0,1/2,1,2$, the contribution to  $g$ is $7.18, 3.95,1.62,0.18$  so graviton emission is an order of magnitude below a naive mode-counting estimate. We ignore angular momentum, which enhances graviton emission \cite{Page:1976ki}.     The physical wavenumber $k$ is  $\tilde{k}/a(t)$, where $\tilde{k}$ is the comoving wavenumber and $a(t)$ is the scale factor.    Integrating,  %
\begin{equation}
\frac{d M_{BH}}{dt} = - \frac{g}{30,720 \pi} \frac{M_p^4}{ M_{BH}^2} \, ,
\end{equation}
from which we can compute the lifetime 
\begin{equation}  \label{eq:lifetime}
\tau   = \frac{10,240 \pi}{ g} \frac{M_{BH}^3}{M_p^4} = \frac{240}{g} \sqrt{\frac{3}{2\pi}} \frac{M_p^5}{\Einit^6} \, .
\end{equation}

An upper bound on $\Einit$ comes from the inflationary energy scale, which is constrained by the non-detection of a primordial gravitational wave background in the CMB, which we (generously) take to be $\sim 10^{16}$GeV.   At the lower end we are interested in black holes which decay prior to nucleosynthesis with time to spare for thermalization, so we need $\tau \lesssim 100$~s. With $\Einit = 10^{12}$~GeV,   $\tau \approx 29/g$~s, and the initial temperature is  $18.8$~TeV.  Standard model states alone give $g\sim {\cal O}(10^2)$ and in what follows we (conservatively) assume $g \ge 10^3$.     Lowering $\Einit$ slightly ensures that the PBH will survive through nucleosynthesis, so we assume  $\Einit \ge 10^{12}$~GeV.

\begin{figure}[tb]
\includegraphics[width=3in]{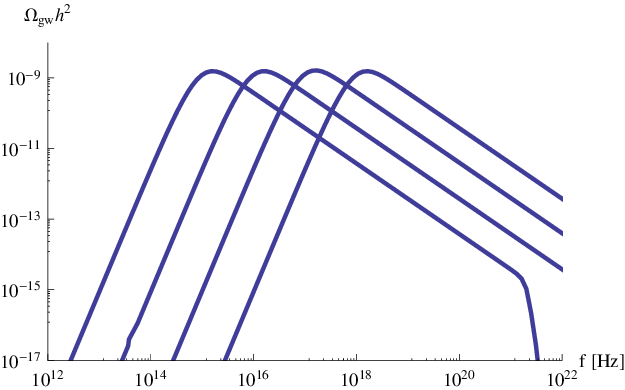}  
\caption{\label{fig:E}   $\Omega_{gw}(f)h^2$ with (from left to right) $\Einit =  10^{15}, 10^{14}, 10^{13}$, and $10^{12}$ GeV, and $\beta=0.001$ and $g=1000$. }
\includegraphics[width=3in]{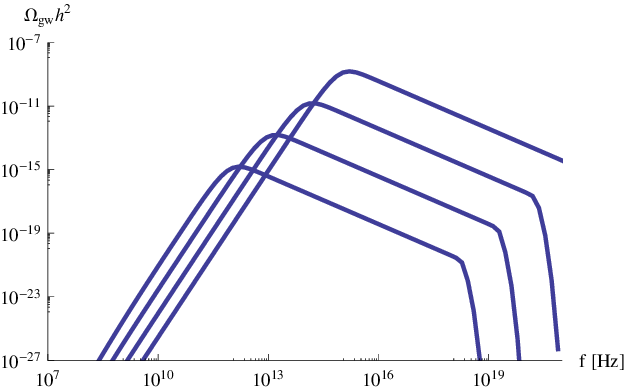}  
\caption{\label{fig:g}    $\Omega_{gw}(f)h^2$ with (from top to bottom) $g= 10^3$, $10^5$, $10^7$ and $10^9$. In all cases $\beta=0.001$ and $\Einit = 10^{15}$GeV. }
 \includegraphics[width=3in]{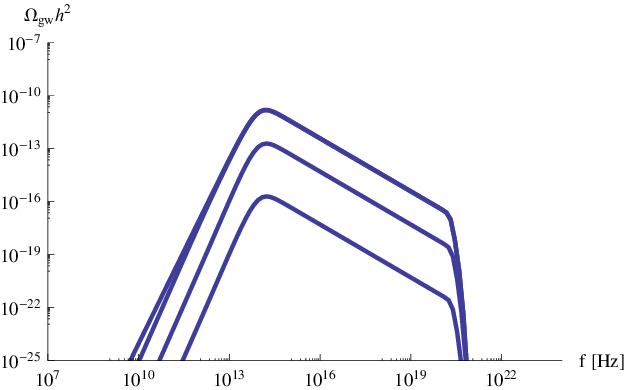}  
\caption{\label{fig:f}  $\Omega_{gw}(f)h^2$ with (from top to bottom) $\beta = 10^{-3}$, $10^{-6}$, $10^{-9}$, and  $10^{-12}$. The $\beta = 10^{-3}$ and $10^{-6}$ cases  lie on top of each other.  In all cases  $g=10^5$ and $\Einit = 10^{15}$GeV.}
\end{figure}

\mbox{}

\noindent {\em Gravitational Wave Background:} 
   Denoting the number density of PBH by $n(t)$, the energy density $\rho_{BH} = n(t) M_{BH}(t)$.   We thus solve\footnote{Gravitons radiated by PBH are considered in \cite{Barrow:1990he,BisnovatyiKogan:2004bk} in a non-expanding universe.  The former omits numerical factors, underestimating the PBH lifetime and  present day frequencies, and the latter focusses on  present-day PBH.   Also,  \cite{Saito:2008jc}  considers gravitational waves generated by the {\em formation\/} of PBH.  }
\begin{eqnarray}
\frac{d \rho_{BH}}{dt} &=& \dot{n}(t) M_{BH} + n(t) \dot{M}_{BH} \, , \\
&=& -3 \frac{\dot{a}}{a} \rho_{BH} + \rho_{BH} \frac{\dot{M}_{BH}}{M_{BH}} \, ,\\
\frac{ d \rho_{rad}}{dt} &=& -4\frac{\dot{a}}{a}\rho_{rad} - \rho_{BH} \frac{\dot{M}_{BH}}{M_{BH}}  \, ,\\
\frac{\dot{a}}{a} &=& \left[\frac{8 \pi }{3 M_p^2} (\rho_{BH}+ \rho_{rad}) \right]^{1/2} \,  
\end{eqnarray}
along with equation~(\ref{eqn:kt}).  One obtains $\Omega_{gw}$ by an appropriate rescaling. We finally compute   the present-day spectral energy density of gravitational radiation \cite{Easther:2006gt,Price:2008hq},
\begin{equation}
\Omega_{gw}(f) = \frac{1}{\rho} \frac{d \rho_{gw}}{d \ln f}
\end{equation}
where $\rho$ is the overall density and $\rho_{gw}$ is the energy density in gravitational waves.\footnote{This quantity depends weakly on $g_\star$, the number of degrees of freedom after the universe rethermalizes. This differs from the $g$ that fixes $\tau$, as a decaying PBH is  much hotter than the surrounding universe. We take $g_\star =200$, and plot $\Omega_{gw}(f)h^2$, where $h$ is the dimensionless Hubble parameter.} 

 Figure~\ref{fig:E}  shows $\Omega_{gw}(f)h^2$ as a function of $\Einit$. The gravitational wave power is substantial, and at very high frequencies.  Roughly speaking, the temperature of the universe scales as $1/a(t)$. A decaying black hole is much hotter than the surrounding universe, but the emitted gravitational waves are redshifted by the same factor as other radiation. Consequently, these gravitational waves have a higher frequency than the present-day CMB. Lowering $\Einit$ increases the PBH lifetime, enhancing this discrepancy and pushing the gravitational wave signal to higher frequencies.  The ``dip" at very high frequencies arises because these quanta can only be sourced by a small black hole, and are produced in smaller numbers.   Conversely, increasing $g$ reduces the fraction of emission into gravitational waves, lowering $\Omega_{gw}$. As $\tau$ is inversely proportional to $g$,  the gravitational waves are emitted when the universe is smaller, increasing the subsequent redshift factor of the emitted radiation, lowering their present day frequency, as seen in Figure~\ref{fig:g}.

 \mbox{}

\begin{figure}[tb]
\includegraphics[width=3in]{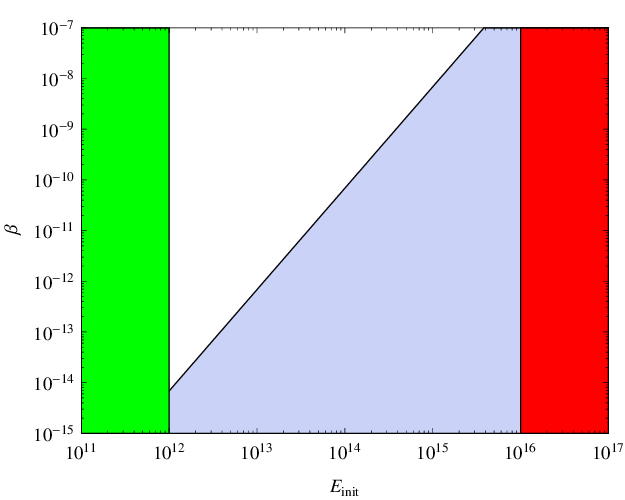}  
\caption{\label{fig:region} The region in the $\{\Einit,\beta\}$ for which a matter dominated period is allowed is plotted for  $g=1000$ (white), along with the generic requirement that  $10^{12}<\Einit<10^{16}$. }
\end{figure}
 
\begin{figure}[tb]
\includegraphics[width=3in]{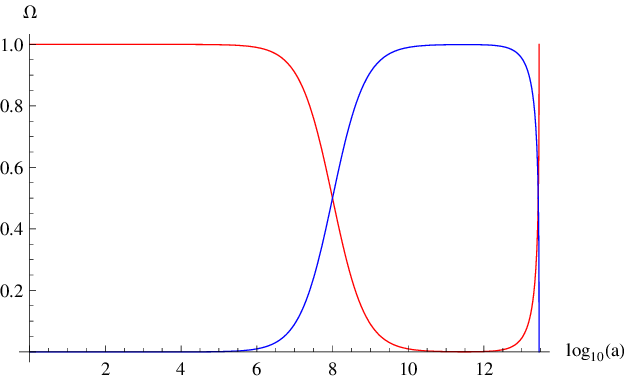}  
\caption{\label{fig:omega}  $\Omega_{BH}$ and $\Omega_{rad}$ are plotted for a scenario with $\Einit = 10^{13}$~GeV, $\beta = 10^{-8}$, and $g=1000$. }
\end{figure}

\noindent {\em Early Matter Domination:}     The primordial universe is   radiation dominated, whereas PBH scale like matter. Initially, $\Omega_{BH} \propto a(t)$ until either $\Omega_{BH} \approx 1 $, or the PBH reach the final phase of their evaporation and $\Omega_{rad}$ begins to grow.  If the universe does become PBH  dominated, all  the radiation in the ``late'' universe will have been emitted by PBH, with the ``original'' radiation making a negligible contribution. In this case $\Omega_{gw} = .36/g$ after evaporation. This is simply the fraction of the total emission in gravitons and $\Omega_{gw}(f)$ is independent of $\beta$. If $\beta$ is very small or $g$ very large, the universe is always radiation dominated and $\Omega_{gw} <.36/g$.    Setting   $\rho_{rad} \approx \Einit^4$,  a matter dominated phase takes place if
\begin{equation} 
\beta \gtrsim \frac{1}{8}  \sqrt{\frac{g}{15}} \frac{\Einit^2}{M_P^2} \, .
\end{equation}
Recall that $H^2 = 1/4t^2$ in a radiation dominated universe. If the universe remains radiation dominated until the PBH have decayed, its grows by $a(\tau+\tinit) \approx (\tau/\tinit)^{1/2}$ with $a(\tinit) \equiv 1$.   Thus,  $\Omega_{gw}(f)$ decreases with $\beta$ if the above inequality is not satisfied, as shown in Figure~\ref{fig:f}.  Figure~\ref{fig:region} shows the region of parameter space for which a matter dominated phase occurs, while Figure~\ref{fig:omega} shows $\Omega_{BH}$ and $\Omega_{rad}$ for a specific scenario with a lengthy matter dominated phase.  

In a radiation dominated universe, $H(t) \sim 1/a(t)^2$.   As always $1/H$ defines the physical Hubble scale while  the {\em comoving\/} Hubble distance is  $a(t)/\Hinit$.  The number of PBH per initial Hubble volume is $\beta$, so before matter domination, the number of PBH per Hubble volume is $\beta a(t)^3$. This number can be large: in Figure~\ref{fig:omega}, $\beta = 10^{-8}$, and $a(t)= 10^8$ before PBH domination, so there are $10^{16}$ PBH within a single Hubble horizon. Perturbations grow in a matter dominated universe.  A mode which is inside the horizon and longer than the Jeans length has amplitude $\delta \propto \eta^2 $, $\eta = \int{dt/a(t)} $   \cite{Peacock:1999ye}. During matter domination, $\delta \sim a(t)$, and short scales become nonlinear.  Moreover, in order to ensure the formation of PBH, the initial amplitude of the perturbations is presumably substantially larger than the canonical $10^{-5}$ found at astrophysical scales.  A  PBH dominated phase may thus be accompanied by the growth of nonlinear structure at sub-horizon scales, leading to the formation of large clusters of PBH.  This situation is reminiscent of the present universe,  with the decaying, clustered PBH playing the role of ``Hawking stars".

The possibility that PBH cause a transient matter dominated phase has been discussed previously (e.g. \cite{Barrow:1990he}) and a universe dominated by decaying PBH is in thermal equilibrium and thus a potential site for baryogenesis \cite{Sakharov:1967dj,Barrow:1990he,Baumann:2007yr}. Crucially, the formation  of nonlinear over-densities could dramatically enhance the interaction rates {\em between\/} black holes \cite{Kotok:1998rp,Chisholm:2005vm}.  If two PBH merge, the resulting black hole lives roughly eight times longer than the parent objects. If a typical PBH survives until shortly before the onset of nucleosynthesis,  a small population of longer lived black holes is potentially troublesome. Since the lifetime of the PBH depends very strongly on the initial energy, we see from equation~(\ref{eq:lifetime}) that a factor of 10 in $\tau$ can be eliminated by increasing $\Einit$ by a factor $10^{1/6} \approx 1.5$ but  the lower bound on $\Einit$ will only change substantially if many PBH coalesce  into single objects. 

To put a crude lower bound on the merger rate, recall that our horizon-mass PBH, have a Schwarzschild radius $r_s = 2M/M_p^2$ which  is equal to the initial Hubble length, $1/H$. Assume that PBH separated by an initial comoving distance of $c r_S$ will merge, where $c$ is a number of order unity. In a comoving region of radius $c r_S$, we expect to find $\sim c^3 \beta$ PBH, so volumes with $N$ PBH will be $\sim (c^3 \beta)^{(N-1)}$ rarer than volumes with just one PBH. Thus, if we reach a matter dominated phase the fraction of the universe composed of PBH with mass $N M_{BH}(\tinit)$, is $\sim (c^3 \beta)^{(N-1)}$.  Unless $\beta$ is close to unity this initial merger phase will not yield a long-lived population of PBH.  However,   correlations in the initial distribution of PBH  \cite{Chisholm:2005vm} or the formation of large,  large nonlinear clusters of PBH could substantially enhance the merger rate. 

\mbox{}

\noindent {\em Discussion:}  We show that light PBH which evaporate before nucleosynthesis lead to a high frequency gravitational wave background.  At present, this is of theoretical interest, given that this  background is at frequencies far beyond the sensitivity region of LIGO, or proposed space-based interferometers such as LISA, which are the most sensitive gravitational wave experiments currently in development. However, the existence of plausible high frequency backgrounds motivates the development of novel detector technologies.  The spectral density of this background is substantial, and may exceed that   obtainable from phase transitions or bubble collisions \cite{Kamionkowski:1993fg,Easther:2006gt}. Further, a light PBH population can lead to a temporary period of matter domination before the onset of nucleosynthesis during which   clusters of PBH could form, leading to a cold phase during which the primordial universe is dominated by clusters of ``Hawking stars".

Gravitational waves generated during preheating or parametric resonance at the end of inflation have recently received considerable attention \cite{Easther:2006gt,Easther:2006vd,Dufaux:2007pt}. Decaying PBH thus provide a further mechanism by which inflation -- if it sources perturbations which lead to the formation of  PBH -- may generate a high frequency gravitational wave background. Very simple models of  inflation do not yield PBH and thus have $\beta \equiv 0$, but  current bounds on the running of the spectral index $\alpha = d n_s/d\ln{k}$ are compatible with PBH production  \cite{Peiris:2008be}. Further,  these bounds are obtained by extrapolating the full inflaton potential  from the region  traversed as astrophysical perturbations are generated. This is  not valid for models where inflation ends abruptly, and these scenarios can lead to substantial PBH production \cite{GarciaBellido:1996qt}, although the simplest models of this form often predict   $n_s>1$, which is in conflict with current data.  Consequently, we simply treat $\beta$ as a free parameter, although it would be computable in any well-specified inflationary scenario. However note that it need not be large -- even if  $\beta< 10^{-10}$ one may still have a lengthy matter dominated phase, provided $\Einit$ is at the lower end of the allowed range. 

The analysis here contains a number of simplifying assumptions. However, these do not affect our basic conclusion, that a high frequency gravitational wave background generated by Hawking radiation is the only signature of a quickly decaying  PBH population which certainly survives until the present epoch.

\section{Acknowledgments}
We thank De-Chang Dai, Andrew Liddle, Daisuke Nagai, Don Page and Dejan Stojkovic for discussions.   RE is supported in part by the United States Department of Energy, grant DE-FG02-92ER-40704 and by an NSF Career Award PHY-0747868.  

\bibliography{pbh}
 
\end{document}